\documentclass[10pt]{article}
\usepackage{amsmath}
\usepackage{tabularx}
\usepackage[utf8]{inputenc}
\usepackage{hyperref}
\usepackage[font=small,labelfont=bf]{caption}
\usepackage{tabularx}
\usepackage[sort,numbers,square]{natbib}
\usepackage{lipsum}
\usepackage{graphicx, color}
\usepackage{hyperref}
\usepackage[version=3]{mhchem}
\usepackage[left=1.5cm, right=1.5cm, top=1.785cm, bottom=2.0cm]{geometry}
\usepackage{balance}
\usepackage{mathptmx}
\usepackage{graphicx} 
\usepackage{lastpage}
\usepackage{float}
\usepackage{fancyhdr}
\usepackage{fnpos}
\usepackage[english]{babel}
\addto{\captionsenglish}{%
  
}
\usepackage{array}
\usepackage{droidsans}
\usepackage{multicol}
\usepackage{charter}
\usepackage[usenames,dvipsnames]{xcolor}
\usepackage{setspace}
\usepackage[compact]{titlesec}
\usepackage{hyperref}
\usepackage{epstopdf}

\definecolor{cream}{RGB}{222,217,201}

\begin{document}

\begin{center}
\LARGE{\textbf{Shear thickening in suspensions of particles with dynamic brush layers}} \\
\vspace{0.2cm}

\Large{Hojin Kim,$^{\ast}$\textit{$^{a,b}$} Michael van der Naald,\textit{$^{a,c}$} Finn A. Braaten,\textit{$^{a,c}$} Thomas A. Witten,\textit{$^{a,c}$} Stuart J. Rowan,\textit{$^{b,d}$} and Heinrich M. Jaeger\textit{$^{a,c}$}}
\vspace{0.2cm}

\noindent
\normalsize{\textit{$^{a}$~James Franck Institute, The University of Chicago, Chicago, Illinois 60637, USA\\
$^{b}$~Pritzker School of Molecular Engineering, The University of Chicago, Chicago, Illinois 60637, USA\\
$^{c}$~Department of Physics, The University of Chicago, Chicago, Illinois 60637, USA\\
$^{d}$~Department of Chemistry, The University of Chicago, Chicago, Illinois 60637, USA\\
$^{\ast}$~hojinkim718@gmail.com}}
\end{center}
 \vspace{0.3cm}

\noindent\normalsize{Control of frictional interactions among liquid-suspended particles has led to tunable, strikingly non-Newtonian rheology via the formation of strong flow constraints as particles come into close proximity under shear. Typically, these frictional interactions have been in the form of physical contact, controllable via particle shape and surface roughness. We investigate a different route, where molecular bridging between nearby particle surfaces generates a controllable `sticky’ friction. This is achieved with surface-functionalized colloidal particles capable of forming dynamic covalent bonds with telechelic polymers that comprise the suspending fluid. At low shear stress this results in particles coated with a uniform polymer brush layer. Beyond an onset stress $\sigma^\ast$ the telechelic polymers become capable of bridging and generate shear thickening. 
Over the size range investigated, we find that the dynamic brush layer leads to dependence of $\sigma^\ast$ on particle diameter that closely follows a power law with exponent -1.76. In the shear thickening regime, we observe an enhanced dilation in measurements of the first normal stress difference $N_1$ and reduction in the extrapolated volume fraction required for jamming, both consistent with an effective particle friction that increases with decreasing particle diameter. These results are discussed in light of predictions for suspensions of hard spheres and of polymer-grafted particles.}

\renewcommand*\rmdefault{bch}\normalfont\upshape
\rmfamily

\vspace{0.5cm}

\section{Introduction}
Shear thickening fluids are materials with a viscosity that rises with increasing shear stress or shear rate \cite{mewis2012colloidal}. Dense suspensions of solid particles suspended in a Newtonian liquid have emerged as the prototypical fluid exhibiting strong shear thickening and have been studied extensively over the last two decades. Understanding the flow of suspensions is critically important in both industrial applications and natural processes. Frictional interaction between particles has been implicated as the dominant contribution to dramatic increases in viscosity during shear thickening, whereby applied stress leads to a growing population of particles interacting frictionally and resisting the applied shear.\cite{seto2013discontinuous,royer2016rheological,lin2015hydrodynamic,hsu2021exploring}

Advances in the understanding of shear thickening in dense suspensions have been facilitated primarily by studying spherical particles as a model system. This is the most straightforward for theoretical calculations and computational modeling as well as being relatively simple to synthesize and measure experimentally. Spherical particle suspensions also have the virtue of making it relatively easy to isolate how changes to particle level characteristics such as particle roughness, particle size, and particle surface chemistry impact shear thickening. For example, by tuning particle roughness \cite{lootens2005dilatant,hsiao2017rheological,bourrianne2022tuning,hsu2018roughness}, varying contact friction \cite{lee2021microstructure}, or incorporating enhanced lubrication hydrodynamics in simulations \cite{jamali2019alternative,wang2020hydrodynamic} it has been shown how the shear thickening response can be controlled. Another way to alter the shear thickening response is to tune the particle size, which controls the stress $\sigma^\ast$ required to make frictional contact and thus to induce shear thickening. For hard sphere suspensions this stress is found to scale as $\sigma^\ast=\frac{F_0}{d^2}$, where the effective applied force $\sigma^\ast d^2$ has to exceed the effective stabilizing force $F_0$ at contact\cite{guy2015towards,krishnamurthy2005shear}.
Whenever local frictional contacts exist between contacting particles, i.e., whenever $\sigma > \sigma^\ast$, relative movement is hindered and this acts as a stress-activated constraint for the associated degree of freedom \cite{guy2015towards, singh2022stress}. 

\begin{figure*}[!t]
    \centering
    \includegraphics[width =0.7\linewidth]{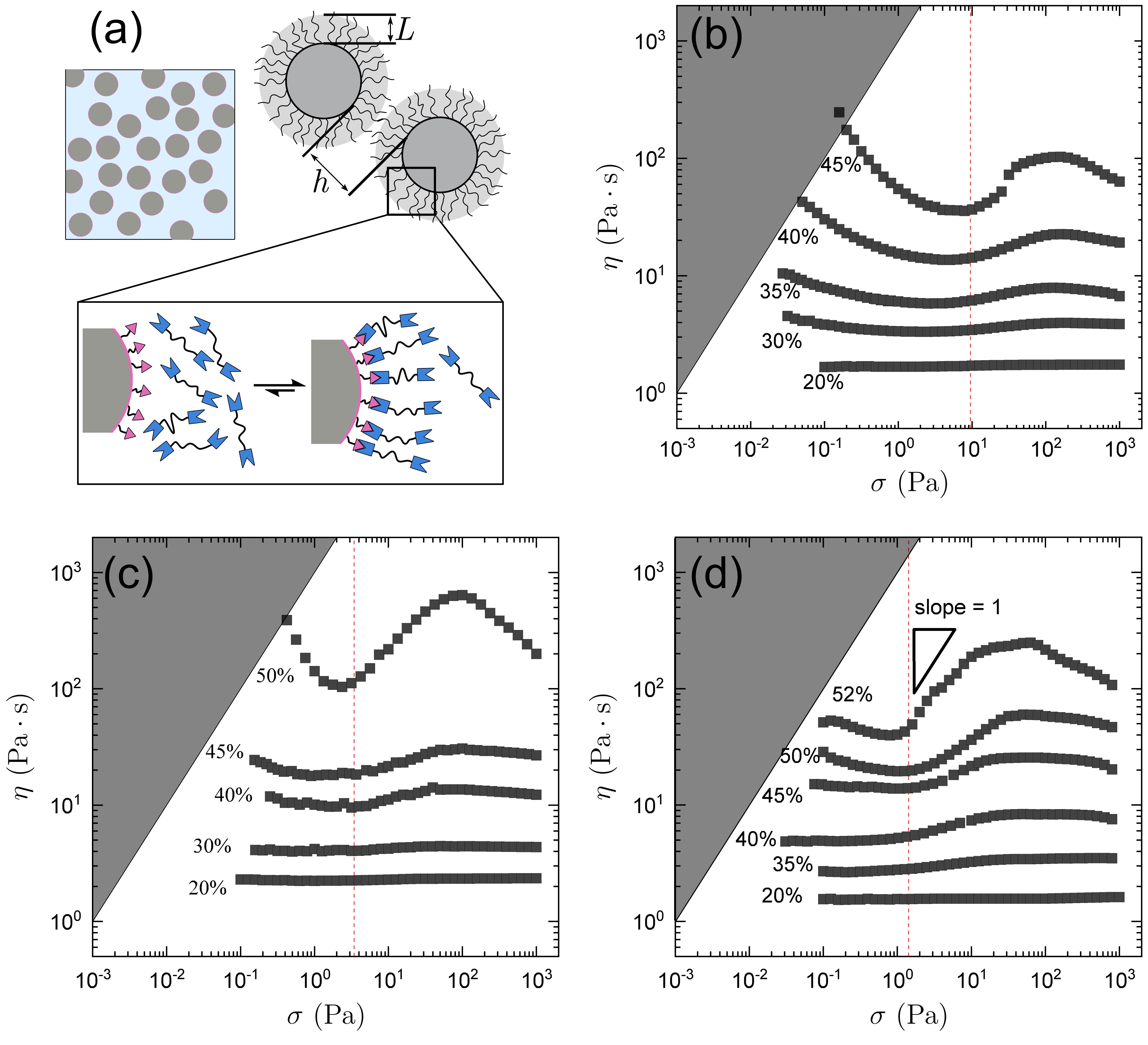}
    \caption{(a) Schematic of silica particles (grey) whose surfaces are functionalized with thiol groups (red triangles) suspended in \textit{p}-nitrobenzalcyanoacetamide (\textit{N}-BCAm)-endcapped poly(propylene glycol) (blue polygons). The polymers form dynamic brushes at the particle surface through the dynamic thia-Michael reaction, which convert to bridging bonds when sheared. (b-d) Viscosity  $\eta$ as a function of applied shear stress $\sigma$ for thiol-functionalized silica particles suspended in \textit{N}-BCAm-endcapped poly(propylene glycol) for different particle volume fraction $\phi$. Particle diameter $d$ is 620nm (SP6, b), 1210nm (SP12, c), and 1930nm (SP19, d). Vertical red dashed lines refer to the average onset stress for shear thickening. Shaded area in left-top corner indicates inaccessible range due to the rheometer minimum rate limit $\dot{\gamma}<0.001$ s\textsuperscript{-1}.}
    \label{fig:phi}
\end{figure*}

Typically, such stress-induced constraints have their origin in physical contact friction. However, friction can also be created by chemical bridging between suitably functionalized surfaces.\cite{raviv2001shear} For example, hydrogen bonding between the particles can enhance the effective friction to the point of inducing shear jamming \cite{james2018interparticle}. However, hydrogen bonds are weak and have a very short lifetime. Comparatively new and less studied are suspensions in which the chemistry of particles and suspending polymeric liquid has been designed to enable bridging via dynamic covalent bonds that can be significantly stronger and have longer lifetimes than non-covalent interactions. These polymers have groups at both ends that bond to the particle surfaces \cite{jackson2022designing,kimdynamic2023}. The effective friction induced by such dynamic covalent bridging has been termed ``sticky friction" because the constraint is generated by solvent molecules that chemically attach to contacting particle surfaces in a reversible manner. These dynamic bonds establish quickly and release more slowly, resulting in a longer lifetime. In addition to strong shear thickening such suspensions also can exhibit anti-thixotopic responses owing to the slow release \cite{jackson2022designing,kimdynamic2023}. The focus of the present work is, however, only on the suspension behavior with increasing applied stress and investigates how dynamic bridging affects the onset of shear thickening, enables a cross-over from lubrication- to friction-dominated rheological behavior, and changes the propensity for jamming. Our experiments show how bridging interactions provide a means of systematically tuning the effective friction in sheared suspensions and, in particular, make it possible to change this effective friction simply by changing particle size.

\section{Experiments}
\subsection{Materials preparation}
Five sets of silica spheres with hydroxyl surface groups were purchased from Fiber Optic Center (New Bedford, MA) with diameters in the range 300 nm $<d<2000$ nm. These particles were surface-functionalized with thiol groups as reported previously (see Supplemental Material) \cite{jackson2022designing}. The size of thiol-functionalized particles was measured as $d=340\pm4$ (SP3), $620\pm10$ (SP6), $870\pm10$ (SP9), $1210\pm40$ (SP12), and $1930\pm90$ nm (SP19) using dynamic light scattering.

The suspending medium for these spheres was a telechelic polymeric liquid comprised of p-nitrobenzalcyanoacetamide (\textit{N}-BCAm)-endcapped poly(propylene glycol) (number-averaged molecular weight, $M_\mathrm{n}\approx5300$ g/mol) \cite{jackson2022designing}. A \textit{dynamic} brush layer is formed on the surface of the particles via the catalyst free, room temperature dynamic thia-Michael reaction between the thiol and the BCAm group. In contrast to a covalently attached brush layer, the dynamic brush continually forms and releases due to the kinetics of the thia-Michael bond at room temperature (Fig.~\ref{fig:phi}a). The equilibrium constant $K_\mathrm{eq}$ of this thia-Michael reaction was measured as $K_\mathrm{eq}\approx 8000$ M\textsuperscript{-1}.\cite{jackson2022designing}
From a thermodynamic perspective, we note that, for all suspensions reported here, the stoichiometric amount of the BCAm Michael-acceptor group is at least 5-fold larger than the surface thiol (1--2 thiol-per-nm\textsuperscript{2}, see the nuclear magnetic resonance spectroscopy study in the Supplemental Material). In this concentration limit and $K_\mathrm{eq}$, the telechelic polymers form dynamic bonds with the surface thiol, resulting in the brush layer. 

\subsection{Rheology measurement}
Suspensions with different volume fractions $\phi$ were prepared by measuring masses of particles and solvents and converting to volumes using their densities. The rheology over all accessible shear stresses, typically $\sigma=10^{-1}$ Pa to $\sigma=10^{3}$ Pa, was measured using a rheometer (Anton Paar, MCR 302 or MCR 702) in cone and plate geometry (25 mm diameter, 2$^\circ$ angle). Owing to the high yield stress, the measurements were performed in rate-controlled mode at low stress prior to the shear thickening regime and otherwise using stress control. For suspensions with the smallest particles, $d=340$ nm, the rheology was measured in rate-controlled mode throughout, given the particularly high yield stress.
The onset stress for shear thickening was determined from the minimum viscosity. 

\section{Results}
Fig.~\ref{fig:phi}b shows the rheology of SP6 at different volume fractions $\phi$. For $\phi=0.30$, the suspension exhibits a mild thinning and thickens slightly at larger stresses. The magnitude of shear thickening increases with $\phi$, as seen in the slope $d\log{\eta}/d\log{\sigma}$. At $\phi=0.45$ the slope reaches 1, which corresponds to discontinuous shear thickening when viscosity is plotted as a function of shear rate. For all $\phi$, the suspension starts to shear thicken at $\sigma^\ast\approx 10$ Pa (red dashed line). 

A second shear-thinning regime appears at high stress. While similar behavior has been found in suspensions of anisotropic\cite{brown2010generality}, irregular\cite{raghavan1997shear}, or soft\cite{jamali2015microstructure} particles, this second thinning in the dynamic suspensions studied here is most likely attributable to the breaking of the bridging interactions at high shear stress.
Probing larger particles, SP12 (Fig.~\ref{fig:phi}c) and SP19 (Fig.~\ref{fig:phi}d), analogous behavior is observed. 
However, the shear thickening behavior for given $\phi$ becomes weaker with increasing $d$. Similarly, the low stress yielding behavior is weaker for larger $d$, even at larger $\phi$.

\begin{figure}
    \centering
    \includegraphics[width =0.6\columnwidth]{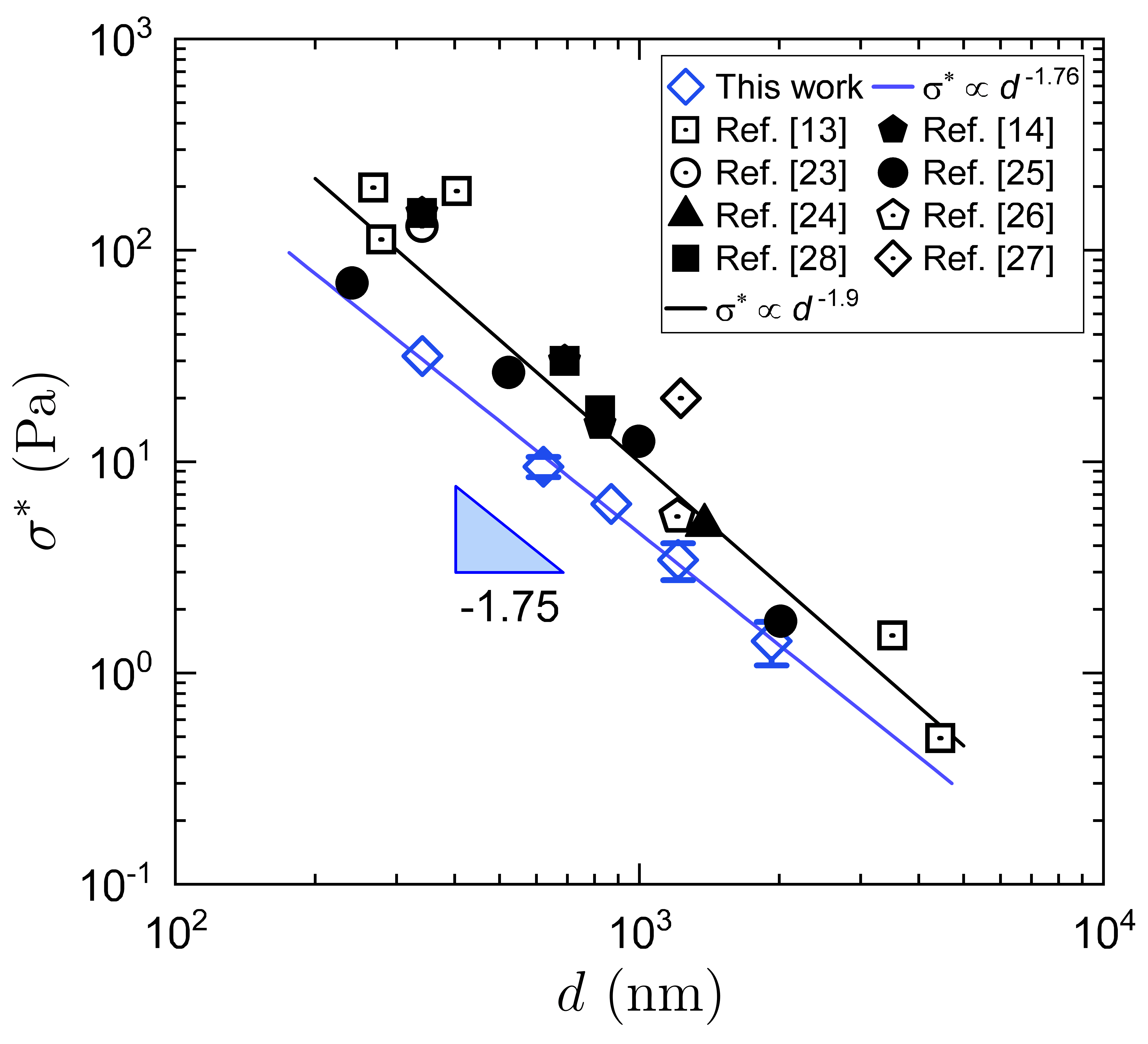}
    \caption{Onset stress for shear thickening $\sigma^\ast$ in suspensions with dynamic covalent bonds (blue diamonds). Also shown are data from prior studies \cite{guy2015towards,mewis2000rheology,frith1996shear,krishnamurthy2005shear,strivens1976shear,smith2010dilatancy,Isa2008,d1993scattering} using suspension with particles having permanently attached covalent brush layers (black symbols). Solid lines are least-squares fits.}
    \label{fig:slope}
\end{figure}

Physically, the onset stress $\sigma^\ast$ signals the stress level at which the particles have been sheared into sufficiently close distance, $h^\ast$, to start forming frictional contact. 
In Fig.~\ref{fig:slope} we plot the onset stress for shear thickening, $\sigma^\ast$, as a function of particle diameter $d$. Over the size range measured, for our particles (blue symbols) this onset closely follows a power law $\sigma^\ast\propto d^\alpha$ with best fit exponent $\alpha = -1.76$.
Also shown in that figure are results from prior work \cite{guy2015towards,mewis2000rheology,frith1996shear,krishnamurthy2005shear,strivens1976shear,smith2010dilatancy,Isa2008,d1993scattering} with particles where the brush layer was attached with permanent covalent bonds (black symbols), and for which a best fit exponent $\alpha \approx -1.9$ has been reported \cite{guy2015towards}. 
We believe that one of the reasons for our data to exhibit very little scatter is that the dynamic brush layers is formed by the exact same \textit{N}-BCAm-endcapped polymers for all particle sizes.

\begin{figure*}
    \centering
    \includegraphics[width =1\textwidth]{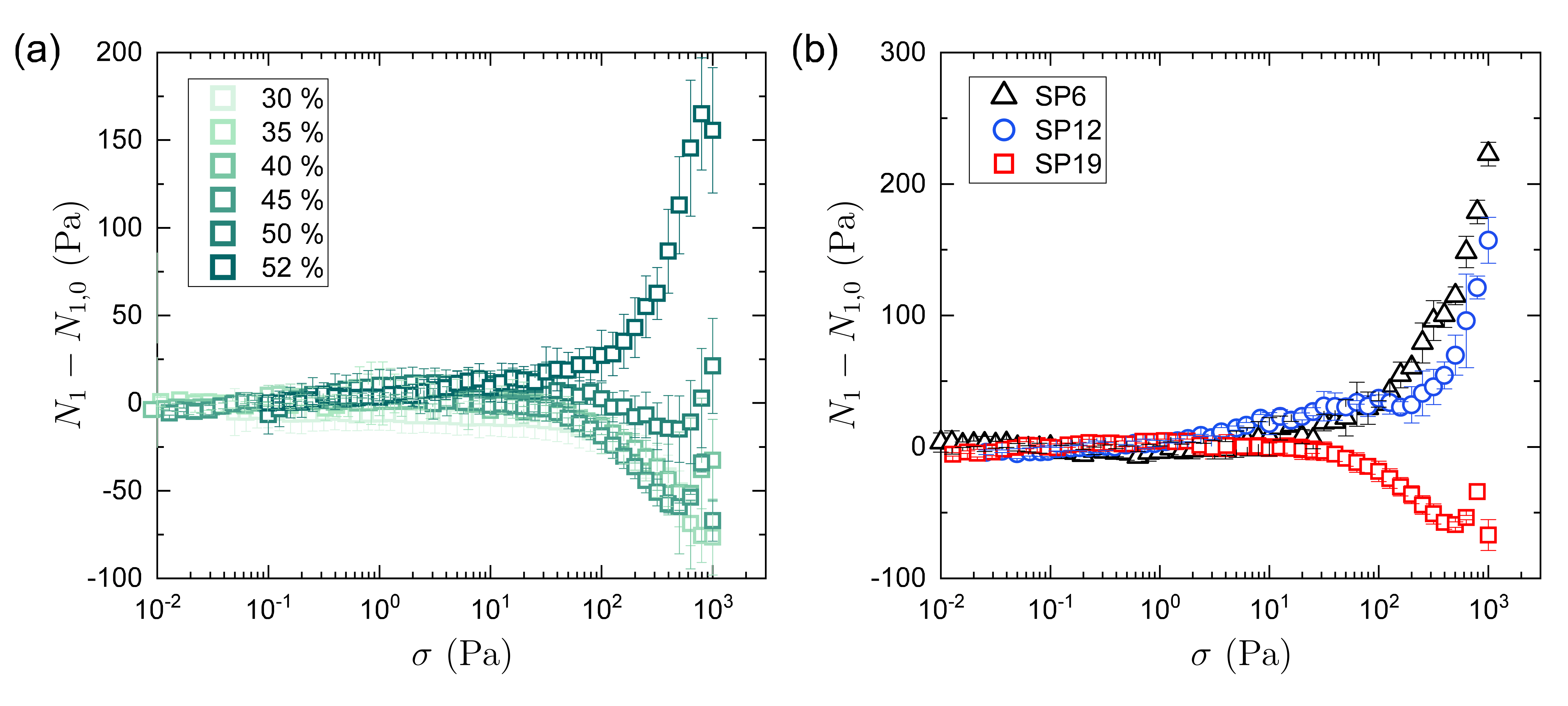}
    \caption{(a) First normal stress difference $N_1$ as a function of shear stress $\sigma$ for suspensions of 1930 nm particles (SP19) with volume fractions $\phi=0.3$ to 0.52. The value measured at the lowest applied shear stress, $N_{1,0}$, has been subtracted. (b) $N_1-N_{1,0}$ is plotted for suspensions of particles with diameter $d=620$ nm (SP6), 1210 nm (SP12), and 1930 nm (SP19) at fixed $\phi=0.45$.}
    \label{fig:N1}
\end{figure*}

As $\sigma$ increases, particles are necessarily pushed closer together, thus promoting bridging by the telechelic macromonomers via dynamic bonds at $\sigma>\sigma^\ast$. This generates constraints to relative particle movement, similar to the constraints arising from a network of frictional contacts. 
In the following, we further explore the dynamic-bond-induced shear thickening state through the first normal stress difference $N_1\equiv\sigma_\mathrm{xx}-\sigma_\mathrm{zz}$. 
Here $\sigma_\mathrm{xx}$ and $\sigma_\mathrm{zz}$ refer to the stress tensor components in the shear flow direction (x) and its gradient direction (z). 
Prior work has suggested that the sign of $N_1$ distinguishes states where particles interact via lubrication ($N_1<0$) or form system-spanning networks of frictional contacts that lead to dilation of the overall sample volume ($N_1>0$) \cite{foss2000structure,park2019contact,royer2016rheological}.
 In Fig.~\ref{fig:N1}a, we plot $N_1 - N_{1,0}$ for the largest particles, 1930nm, at different volume fractions $0.3\leq\phi\leq0.52$, where the initial value $N_{1,0}$ at the lowest applied shear stress has been subtracted. 
For $\phi\leq0.5$, we find $N_1<0$ even beyond the stress $\sigma^\ast\approx1.3$ Pa that indicates the onset of shear thickening. 
We interpret this as indicating that system-spanning networks of frictional contacts can only form for $\sigma > \sigma^\ast$. 
This changes once $\phi \geq 0.52$ where now the sign of $N_1$ reverses from negative to positive as $\sigma\geq\sigma^\ast$. 

The effect on $N_1$ of changing particle size is shown in Fig.~\ref{fig:N1}b for fixed $\phi=0.45$. 
Strikingly, the sign of $N_1$ in the shear thickening regime reverses from negative to positive for particles with $d<1210$ nm (see Supplemental Material for other particle packing fractions). 

\begin{figure}
    \centering
    \includegraphics[width =0.6\columnwidth]{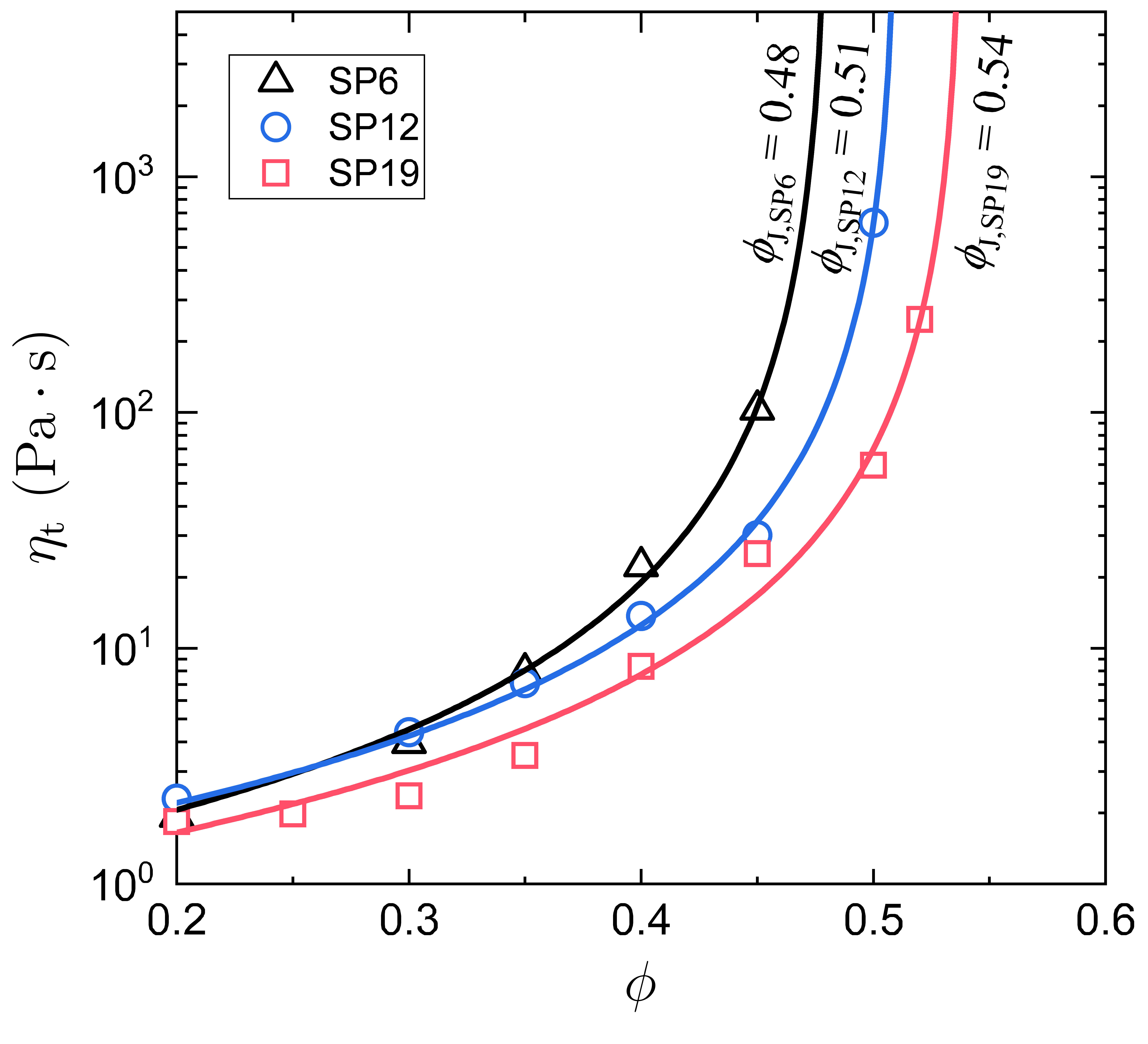}
    \caption{\label{fig:phi_m}Viscosity $\eta_\mathrm{t}$ at the thickened state versus volume fraction $\phi$ for SP6 (black), SP12 (blue), and SP19 (red). Lines are least-squares fits to $\eta=A(1-\phi/\phi_\mathrm{J})^{-n}$. The jamming volume fraction $\phi_\mathrm{J}$ are $\phi_\mathrm{J,SP6}=0.48$, $\phi_\mathrm{J,SP12}=0.51$, and $\phi_\mathrm{J,SP19}=0.54$.}
\end{figure}

The size-dependent magnitude of the effective friction seen in the behavior of $N_1$ is also observed in the approach to jamming as the particle volume fraction $\phi$ is increased. 
This can be seen in plots of the high-stress viscosity $\eta_\mathrm{t}$, measured at the upper end of the shear thickened state, as a function of particle volume fraction $\phi$, which diverges at a friction-dependent jamming volume fraction $\phi^{\mu}_\mathrm{J}$ \cite{mari2014shear}.
The $\phi^{\mu}_\mathrm{J}$ divergence shifts to a lower value with increased interparticle friction and can therefore be taken as another indicator of the effect of `sticky’ friction due to dynamic bridging. 
In Fig.~\ref{fig:phi_m} we plot data for the three particle systems measured and fit them to a power-law relation, $\eta=A(1-\phi/\phi_\mathrm{J})^{-n}$ \cite{andreotti2012shear,lerner2012unified,mari2014shear} to extract $\phi^{\mu}_\mathrm{J}$. 
The fits, with $n$ = 1.7-1.8 (see Supplemental Material), give $\phi^{\mu}_\mathrm{J}$ = 0.54, 0.51, and 0.48 for SP19, SP12 and SP6, respectively.

\section{Discussion}
For particles coated with a polymer brush layer we can expect that shearing them into contact produces a balance between the applied stress and the counteracting stress from the stiffness of the brush.
Balancing the force $\sigma^\ast d^2$ with the restoring force derived from gradient of the interparticle potential for brush-layer stabilized particles \cite{krishnamurthy2005shear} gives
\begin{equation}\label{eqn:criStress}
\sigma^\ast d^2=-\left.\frac{\partial U(r)}{\partial r}\right|_{r=d+h^\ast}.
\end{equation}

Different interparticle potentials will produce different scalings $\sigma^\ast \propto d^{\alpha}$ of the onset stress as a function of particle diameter $d$. For example, stabilization by steric interactions with distance-independent forces
 $\frac{\partial U(r)}{\partial r}$ = const. leads to $\sigma^\ast\propto d^{-2}$. 
The data shown Fig.~\ref{fig:slope} by the black symbols therefore has been interpreted as the result of behavior close to the hard sphere limit.
Another much-studied limit is that for compressed semidilute polymer brush layers\cite{fredrickson1991drainage,krishnamurthy2005shear} with polymer brush thickness $L$ and surface-to-surface distance $h$ (see Fig.~\ref{fig:phi}a).
In this case the stabilizing potential gradient between polymer brush layers was found consistent with $-\frac{\partial U(r)}{\partial r}\sim d^{0.25}$. This gradient gives a scaling $\sigma^\ast\propto d^{-1.75}$\cite{krishnamurthy2005shear}. 
Intriguingly, this closely matches our data for dynamic brushes.
Notably, however, the power law exponent -0.25 in this scaling argument requires semidilute, good solvent conditions for the brush, which are difficult to justify in our system, where we have a very different situation: the particles are dispersed in a melt of \textit{N}-BCAm endcapped polymers and these polymers also form the dynamic brush layer.

The size dependence of the sign reversal in $N_1$ (Fig.~\ref{fig:N1}b) suggests that the strength of the effective friction produced by the bridging linkers increases with decreasing $d$. 
To rationalize this, we consider how particle-particle interactions due to polymer linkages depend on the total energy density associated with bridging. 
This energy density ($\rho_\mathrm{B}$) depends on the energy per bridging bond $\epsilon_\mathrm{B}$, the interparticle contact area $A\sim Ld$ for $d\gg L$, and the number of particles per unit volume, which is proportional to $d^{-3}$, and it scales as $\rho_\mathrm{B}\sim\epsilon_\mathrm{B}Ld/d^3 =\epsilon_\mathrm{B}Ld^{-2}$. 
As a result, although telechelic polymers on larger particles with less curved surfaces have a stronger attractive pair potential, particularly in the regime of $d\gg R_\mathrm{g}$ \cite{meng2006telechelic}, the overall bridging energy density scales inversely with particle size, $\rho_\mathrm{B}\sim d^{-2}$.
Hence, per unit volume smaller particles are more strongly interacting by bridging and behave more frictional, i.e., they experience stronger constraints to relative motion under shear. 
This then decreases the volume fraction required for the formation of system-spanning frictional networks and thus for the observation of behavior with $N_1>0$. 
We also see from Fig.~\ref{fig:N1} that in suspensions where $N_1 < 0$ initially, there can be a sign reversal at larger shear stresses.
We take this as an indication that telechelic linkers make it possible to establish extended network structures even under large applied shear.
However, the decrease in viscosity seen in Fig.~\ref{fig:phi} at these large stress levels suggests that such networks structures deform more easily under shear than hard spheres would.

The fits to the data in Fig.~\ref{fig:phi_m} make it possible to estimate the associated effective friction coefficient $\mu$, using the simulation data by Singh \textit{et al.}\cite{singh2018constitutive}.
Taking $\phi_\mathrm{J}^\infty\approx0.47$ as obtained for infinite sliding friction $\mu_\mathrm{s}=\infty$ at fixed rolling friction $\mu_\mathrm{r}=0.3$\cite{singh2022stress}, the resulting values are $\mu$ = 3.5, 0.9, and 0.5 for SP6, SP12, and SP19, respectively.
The finding that the extracted effective friction coefficient can be larger than unity again highlights that we are dealing with `sticky' friction due to chemical bridging, as opposed to physical contact friction for which $\mu\leq 1$ can typically be expected.
We note that it is possible to attribute some of the decrease in $\phi^{\mu}_\mathrm{J}$ with particle diameter $d$ also to the brush layer effectively increasing the volume fraction of particles in the suspension.
Accounting for the brush layer would then shift $\phi$ to $\phi_\mathrm{eff}=\phi (1+2L/d)^3$. However, this shift is too small to fully explain the observed particle size dependence in $\phi^{\mu}_\mathrm{J}$ (see Supplemental Material).

\section{Summary and Conclusions}
We investigated shear thickening for a class of dense suspensions with particles designed to interact through chemical bridging interactions. 
In these systems silica particles are surface-functionalized with thiol groups and suspended in a \textit{N}-BCAm-endcapped poly(propylene glycol). 
This combination of particle surface and solvent chemistry allows for a dynamic brush layer to be established at the particle surface.
At low shear this brush layer stabilizes the particles. 
As larger stress is applied to the suspension, particles come into sufficiently close proximity that dynamic bridging interactions are formed.
These bridges introduce an effective interparticle friction and promote shear thickening. 
We find that the onset stress of shear thickening for suspensions with dynamic brush layers scales with particle size as $d^{-1.76}$.
It is intriguing that this lines up closely with the theoretical model for covalently-attached brush stabilized particles, which predicts $d^{-1.75}$ but applies to a rather different regime.
The degree to which this model can be generalized is currently unknown.
Our results show that bridging interactions can give rise to dilatancy and form system-spanning networks, just as in shear thickening suspensions where the particles interact sterically through direct contact friction. 
However, owing to the increase in the overall area of contact for fixed brush thickness, and in contrast to steric interactions, the effective bridging-induced friction increases as particle size decreases.
Consequently, also the onset of jamming is shifted to smaller $\phi^{\mu}_\mathrm{J}$ for smaller particles. 

Our findings demonstrate that dynamic bridging interactions provide a direct, highly tunable means for controlling the effective particle friction. For fixed particle and solvent chemistry, this friction can be changed straightforwardly by changing the particle size and it can become as large as or even exceed physical friction produced by particle surface roughness. 
While we reported on one specific dynamic covalent chemistry, the results introduce a versatile way to design the non-Newtonian rheology of shear thickening fluids, and we expect that this approach can be extended further by tuning the molecular structure of the suspending polymer medium, including the molecular weight (length scale of chemical `sticky’ friction), persistence length (rolling friction), and thermodynamic reaction equilibrium (effective friction lifetime).

\section*{Author Contributions}
HK, SJR, and HMJ conceived the research project; HK and FAB performed the experiments; HK, MvdN, TAW, SJR, and HMJ analyzed the data; All authors participated in the discussion and writing the manuscript.

\section*{Conflicts of interest}
There are no conflicts to declare.

\section*{Acknowledgements}
We thank Norman Wagner for fruitful discussion.
The authors acknowledge support from the University of Chicago Materials Research Science and Engineering Center (MRSEC), which is supported by the National Science Foundation under award DMR-2011854. This work was partially supported by National Science Foundation under award DMR-2104694. Parts of this work were carried out at the Soft Matter Characterization Facility and at the MRSEC Characterization Facility at the University of Chicago (award number DMR-2011854).

\graphicspath{{fig/}}

\renewcommand{\thetable}{S\arabic{table}}
\renewcommand{\thefigure}{S\arabic{figure}}
\renewcommand{\theequation}{S\arabic{equation}}
\renewcommand{\thesection}{S\arabic{section}.}
\setcounter{figure}{0}
\setcounter{section}{0}
\date{
	$^a$James Franck Institute, The University of Chicago, Chicago, Illinois 60637, USA\\
	$^b$Pritzker School of Molecular Engineering, The University of Chicago, Chicago, Illinois 60637, USA\\
 $^c$Department of Physics, The University of Chicago, Chicago, Illinois 60637, USA\\
    $^d$ Department of Chemistry, The University of Chicago, Chicago, Illinois 60637, USA\\ 
    $^\ast$ hojinkim718@gmail.com\\
    [2ex]
}
\title{Electronic supplementary information:\\
Shear thickening in suspensions of particles with dynamic brush layers}
\small{\author{Hojin Kim$^{\ast a,b}$, Michael van der Naald$^{a,c}$, Finn A. Braaten$^{a,c}$,\\ Thomas A. Witten$^{a,c}$, Stuart J. Rowan$^{b,d}$, and Heinrich M. Jaeger$^{a,c}$}}

\maketitle

\section{Particle synthesis and characterization}
Silica particles were purchased from Fiber Optic Center (New Bedford, MA) with diameters 300 nm, 500 nm, 1 $\mathrm{\mu m}$, and 2 $\mathrm{\mu m}$. The particle surface was functionalized by the procedure reported previously\cite{crucho2017functional}. First, 10 g of silica particles were added to 400 mL of toluene. The suspension was sonicated for 1 hr and stirred for another 30 min to homogenize the suspension. Then 3-mercaptopropyl trimethoxysilane (MPTMS) was added to the solution. The concentration of MPTMS was kept 10 MPTMS molecules per nm\textsuperscript{2} surface area of the added silica particles. The solution was then heated to reflux and left for 24 hr. Toluene from the final solution was completely removed using a rotary evaporator. The particle powder was washed with ethanol three times by repeating sonication and centrifugation. The final particles were left under vacuum for 24 hr.

\begin{table}
\caption{\label{tab:SH} Estimated thiol density of the particle surface.}
\begin{tabularx}{1\textwidth}{ 
  | >{\centering\arraybackslash}X 
  | >{\centering\arraybackslash}X 
  | >{\centering\arraybackslash}X 
  | >{\centering\arraybackslash}X 
  | >{\centering\arraybackslash}X 
  | >{\centering\arraybackslash}X 
  |}
 \hline
 diameter & 340 nm & 620 nm & 870 nm & 1210 nm & 1930 nm \\
 \hline
 Thiol density of the particle surface & 1.09 SH/nm\textsuperscript{2}  & 2.15 SH/nm\textsuperscript{2}  & 2.19 SH/nm\textsuperscript{2} & 1.20 SH/nm\textsuperscript{2} & 1.31 SH/nm\textsuperscript{2} \\
\hline
\end{tabularx}\centering
\end{table}

\begin{figure}
    \centering
    \includegraphics[width =0.5\linewidth]{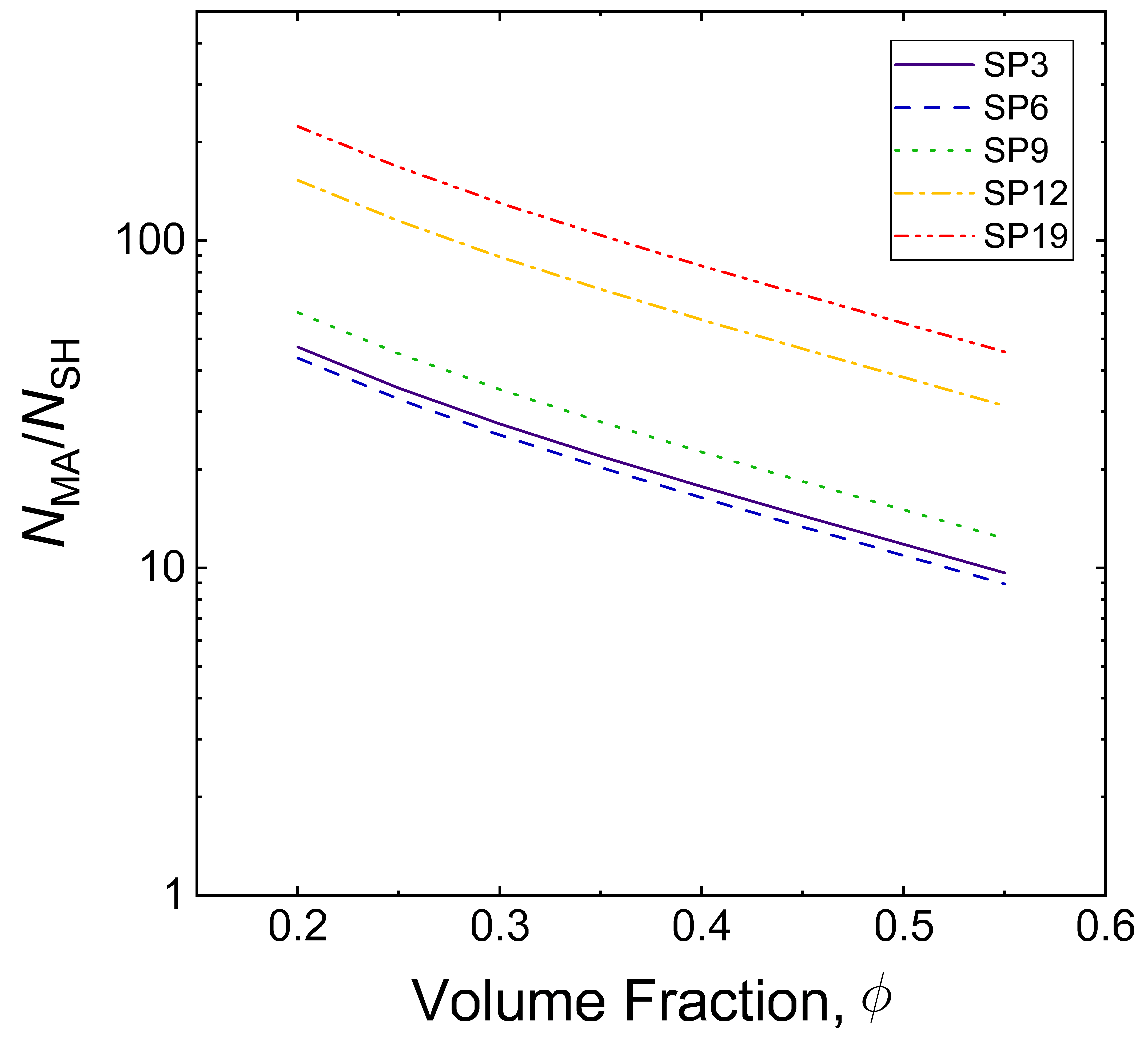}
    \caption{\small{Stoichiometric ratio of Michael-acceptor to thiol in suspensions of different particle sizes is plotted over the particle volume fraction range studied in this study.}}
    \label{fig:ratio}
\end{figure}

The surface density of thiol functional groups was estimated using nuclear magnetic resonance (NMR) studies\cite{crucho2017functional} using 1,3,5-trioxane as an internal standard. First, 20 mg of the silica particles was added to 0.5 M NaOD/D\textsubscript{2}O solution. The solution was stirred at 85 $^\circ$C overnight to fully dissolve the silica particle. The thiol density of the particle surface $\rho_\mathrm{SH}$ was estimated by NMR. The peak intensity
of the (ONa)3SiCH2CH2C\textbf{H}2SH ($\delta= 2.43$ ppm) was compared to that of the internal standard (Table~\ref{tab:SH}). The total number of the surface thiol $N_\mathrm{SH}$ at the particle with diameter $d$ in a suspension with packing fraction $\phi$ is estimated by 
\[N_\mathrm{SH}=\phi\frac{\pi d^2}{\pi d^3/6}\rho_\mathrm{SH},\]
and the total number of the Michael-acceptor group is estimated by
\[ N_\mathrm{MA}=(1-\phi)\rho_\mathrm{PPG}N_\mathrm{A}n_\mathrm{MA}/M_\mathrm{PPG}\]
with the Avogadro constant $N_\mathrm{A}$, the density ($\rho_\mathrm{PPG}\approx1$ g/ml) and molecular weight ($M_\mathrm{PPG}=5300$ g/mol) of (N-BCAm)-endcapped poly(propylene glycol), and the number of Michael-acceptor group $n_\mathrm{MA}=2$.
The ratio of Michael-acceptor to thiol $N_\mathrm{MA}/N_\mathrm{SH}$ in suspensions with various particle volume fractions is plotted in Figure~\ref{fig:ratio}.

\begin{table}
\caption{\label{tab:phi_m}Parameters from a power-law relation, $\eta=A(1-\phi/\phi_\mathrm{J})^{-n}$.}
\begin{tabularx}{0.8\textwidth}
{ 
  | >{\centering\arraybackslash}X 
  | >{\centering\arraybackslash}X 
  | >{\centering\arraybackslash}X 
  | >{\centering\arraybackslash}X | }
 \hline
 diameter & $A$ & $n$ & $\phi_\mathrm{J}$\\
\hline
 620 nm  & 0.79  & 1.79 & 0.48 \\
\hline
 1210 nm  & 0.96  & 1.69 & 0.51 \\
 \hline
 1930 nm  & 0.74  & 1.73 & 0.54 \\
 \hline
\end{tabularx}\centering
\end{table}

\begin{figure}
    \centering
    \includegraphics[width =0.5\linewidth]{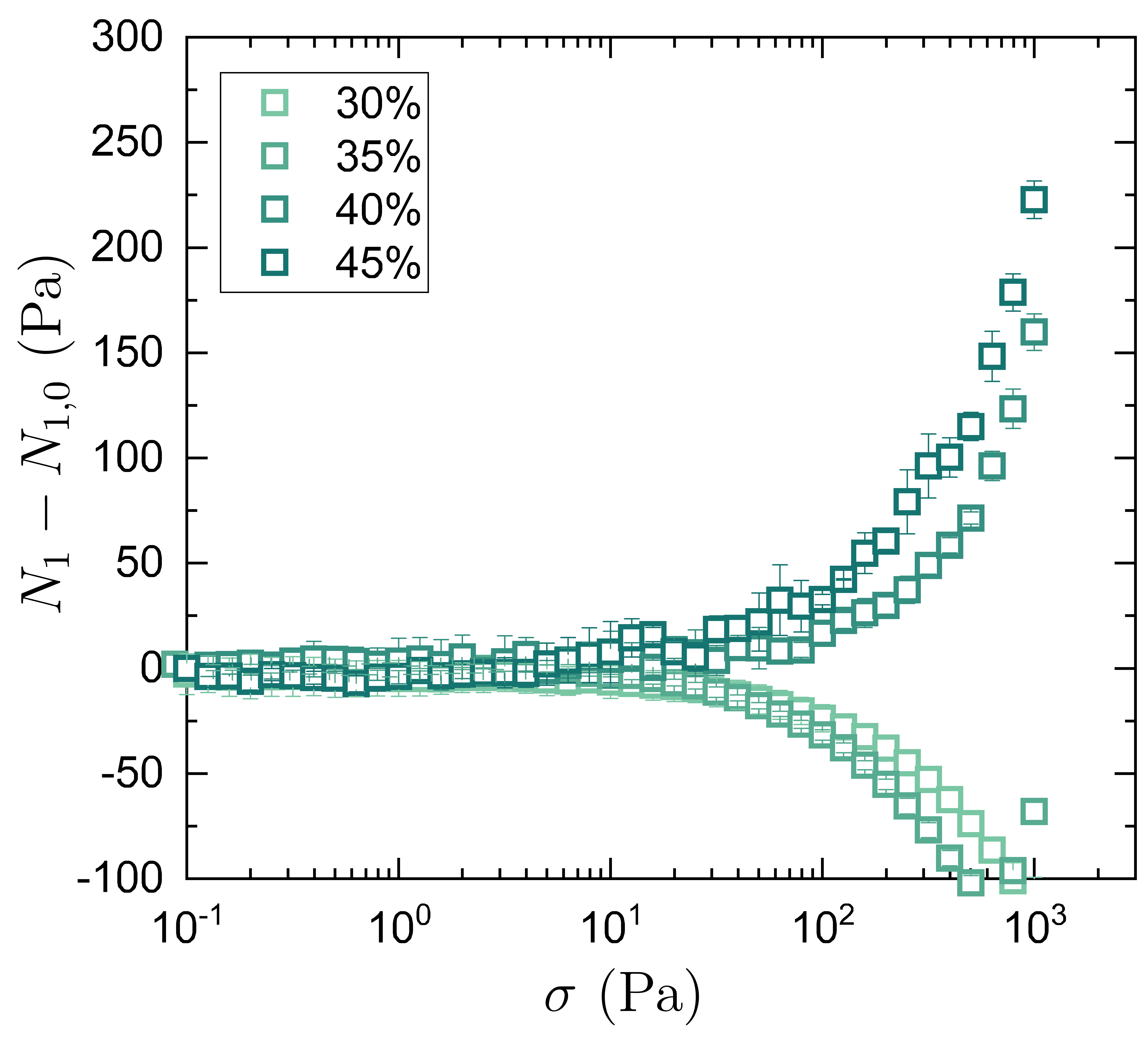}
    \caption{\small{First normal stress difference $N_1$ for particles ($d=620$ nm, SP6).}}
    \label{fig:SP6_N1}
\end{figure}

\begin{figure}
    \centering
    \includegraphics[width =0.5\linewidth]{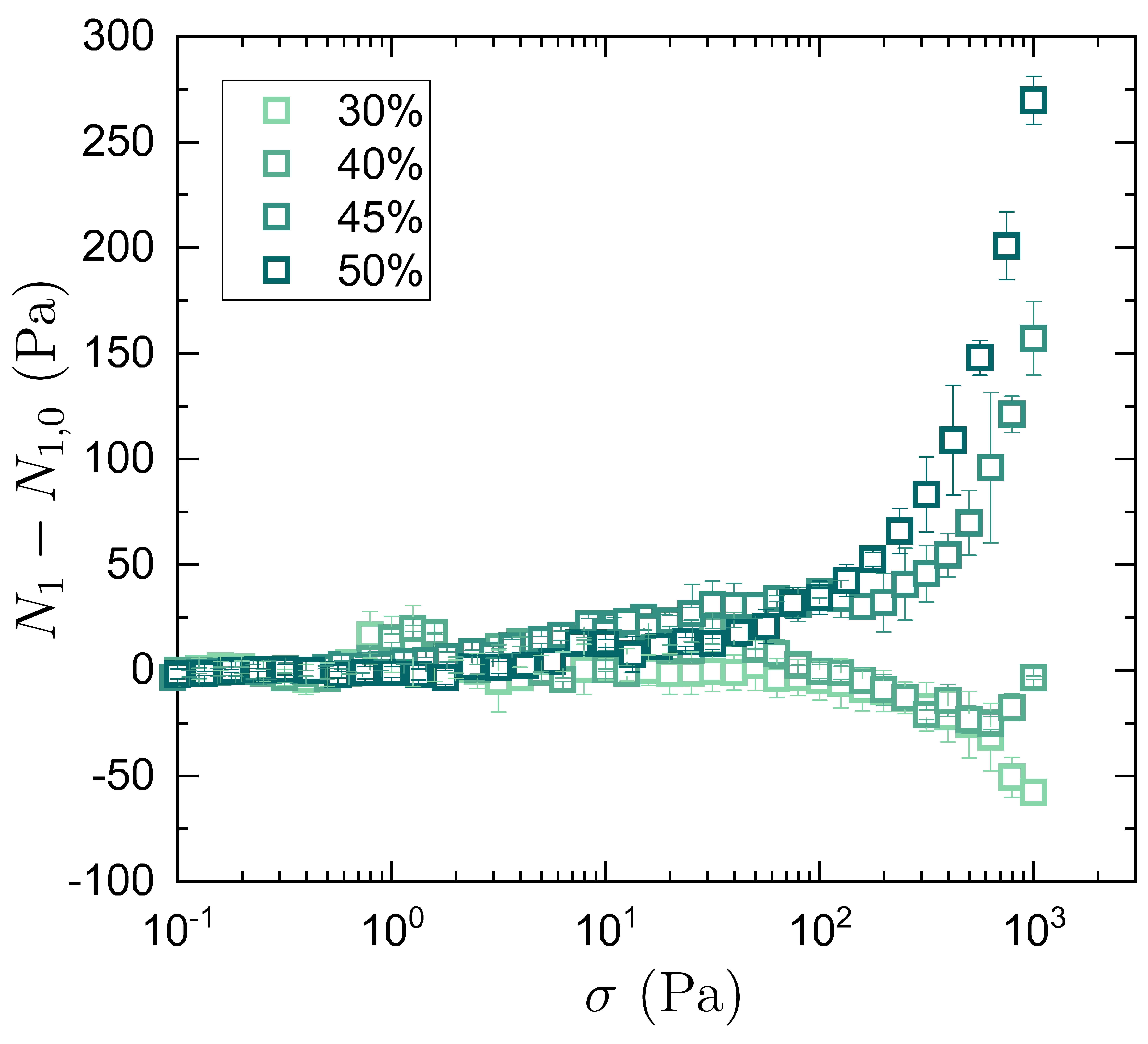}
    \caption{\small{First normal stress difference $N_1$ for particles ($d=1210$ nm, SP12).}}
    \label{fig:1um_N1}
\end{figure}

\begin{figure}
    \centering
    \includegraphics[width =0.9\linewidth]{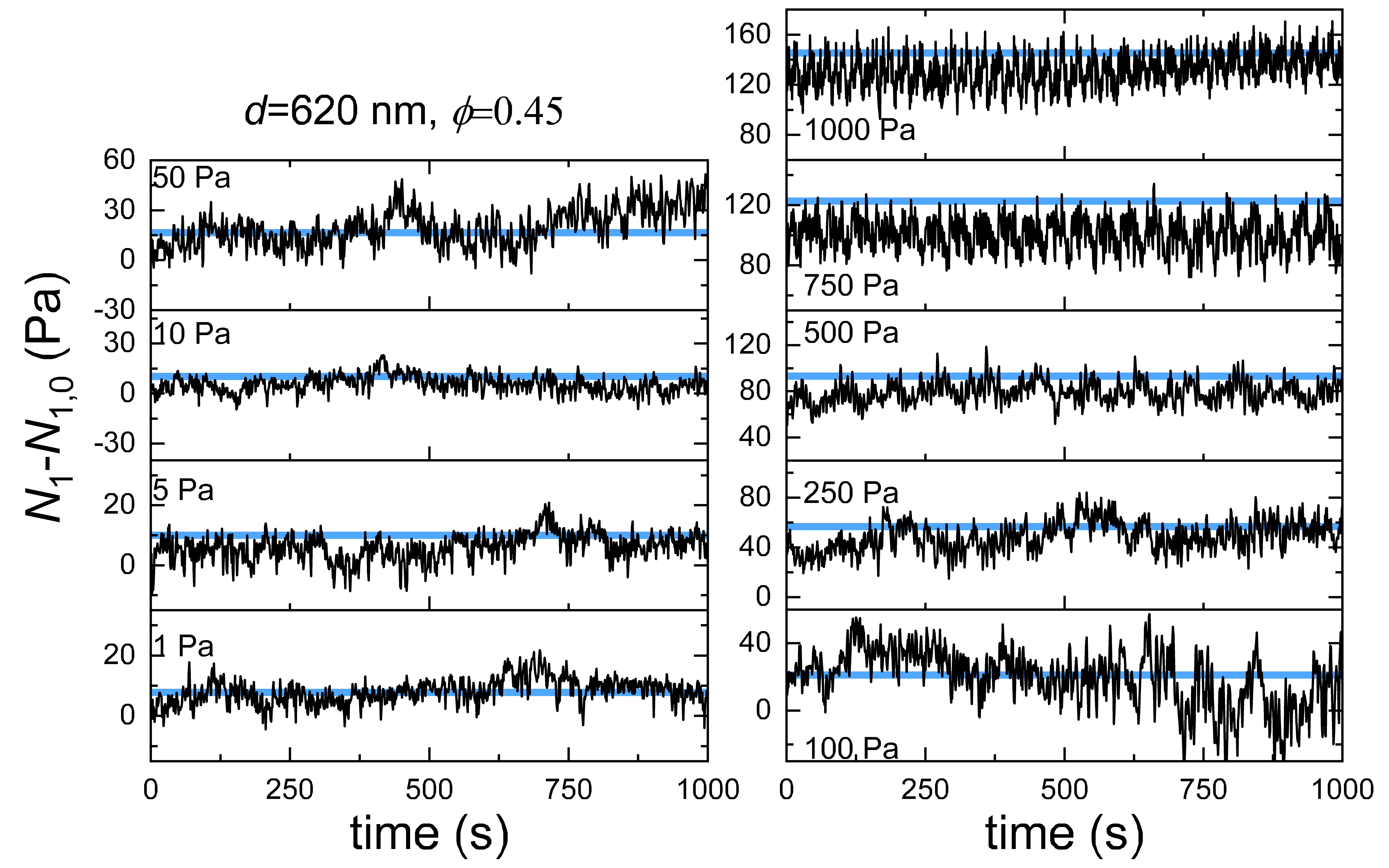}
    \caption{\small{Time-dependent measurement of the first normal stress difference $N_1$ for the suspension of particles with diameter $d=620$ nm at $\phi=0.45$. Measurements were performed at each shear stress after the consistent preshear and relaxation steps. Blue lines indicate the $N_1$ at each shear stress from the stress-sweep measurement (30 s per each stress).}}
    \label{fig:time}
\end{figure}

\begin{figure}
    \centering
    \includegraphics[width =0.5\linewidth]{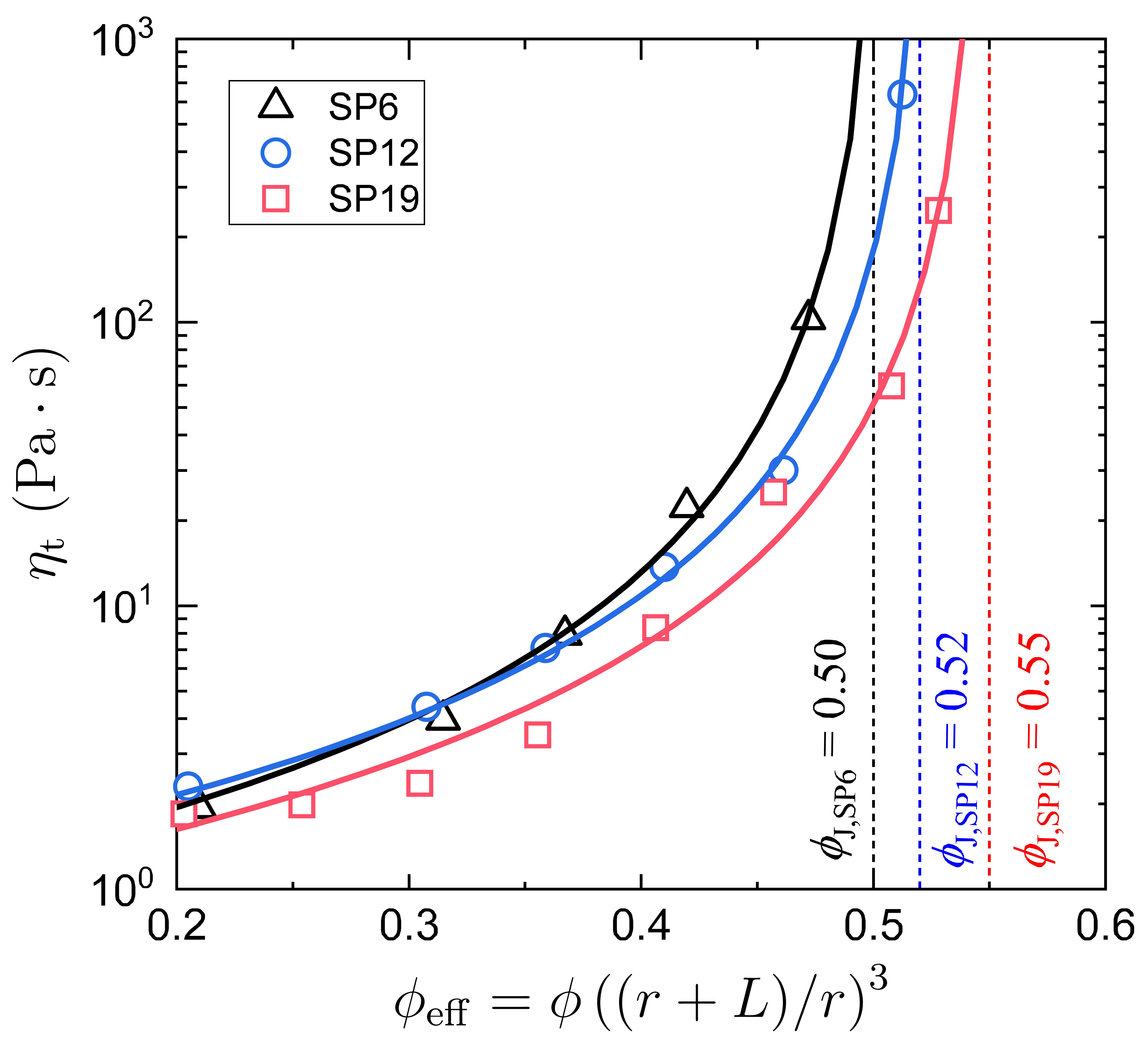}
    \caption{\small{Viscosity $\eta_\mathrm{t}$ at the thickened state versus the effective volume fraction $\phi_\mathrm{eff}=\phi  (1+2L/d)^3$ for SP6 (black), SP12 (blue), and SP19 (red). The thickness of the brush layer is set to half of the contour length $L_\mathrm{c}\approx 10$ nm.
    Lines are least-squares fits to $\eta=A(1-\phi_\mathrm{eff}/\phi_\mathrm{J})^{-n}$. The estimated jamming volume fraction $\phi_\mathrm{J}$ is $\phi_\mathrm{J,SP6}=0.50$, $\phi_\mathrm{J,SP12}=0.52$, and $\phi_\mathrm{J,SP19}=0.55$.}}
    \label{fig:phi_m_eff}
\end{figure}


\begin{thebibliography}{10}

\bibitem{mewis2012colloidal}
Jan Mewis and Norman~J Wagner.
\newblock {\em Colloidal suspension rheology}.
\newblock Cambridge University Press, 2012.

\bibitem{seto2013discontinuous}
Ryohei Seto, Romain Mari, Jeffrey~F Morris, and Morton~M Denn.
\newblock Discontinuous shear thickening of frictional hard-sphere suspensions.
\newblock {\em Physical Review Letters}, 111(21):218301, 2013.

\bibitem{royer2016rheological}
John~R Royer, Daniel~L Blair, and Steven~D Hudson.
\newblock Rheological signature of frictional interactions in shear thickening suspensions.
\newblock {\em Physical Review Letters}, 116(18):188301, 2016.

\bibitem{lin2015hydrodynamic}
Neil~YC Lin, Ben~M Guy, Michiel Hermes, Chris Ness, Jin Sun, Wilson~CK Poon, and Itai Cohen.
\newblock Hydrodynamic and contact contributions to continuous shear thickening in colloidal suspensions.
\newblock {\em Physical Review Letters}, 115(22):228304, 2015.

\bibitem{hsu2021exploring}
Chiao-Peng Hsu, Joydeb Mandal, Shivaprakash~N Ramakrishna, Nicholas~D Spencer, and Lucio Isa.
\newblock Exploring the roles of roughness, friction and adhesion in discontinuous shear thickening by means of thermo-responsive particles.
\newblock {\em Nature Communications}, 12(1):1477, 2021.

\bibitem{lootens2005dilatant}
Didier Lootens, Henri Van~Damme, Yacine H{\'e}mar, and Pascal H{\'e}braud.
\newblock Dilatant flow of concentrated suspensions of rough particles.
\newblock {\em Physical Review Letters}, 95(26):268302, 2005.

\bibitem{hsiao2017rheological}
Lilian~C Hsiao, Safa Jamali, Emmanouil Glynos, Peter~F Green, Ronald~G Larson, and Michael~J Solomon.
\newblock Rheological state diagrams for rough colloids in shear flow.
\newblock {\em Physical Review Letters}, 119(15):158001, 2017.

\bibitem{bourrianne2022tuning}
Philippe Bourrianne, Vincent Niggel, Gatien Polly, Thibaut Divoux, and Gareth~H McKinley.
\newblock Tuning the shear thickening of suspensions through surface roughness and physico-chemical interactions.
\newblock {\em Physical Review Research}, 4(3):033062, 2022.

\bibitem{hsu2018roughness}
Chiao-Peng Hsu, Shivaprakash~N Ramakrishna, Michele Zanini, Nicholas~D Spencer, and Lucio Isa.
\newblock Roughness-dependent tribology effects on discontinuous shear thickening.
\newblock {\em Proceedings of the National Academy of Sciences}, 115(20):5117--5122, 2018.

\bibitem{lee2021microstructure}
Yu-Fan Lee, Yimin Luo, Tianyi Bai, Carlos Velez, Scott~C Brown, and Norman~J Wagner.
\newblock Microstructure and rheology of shear-thickening colloidal suspensions with varying interparticle friction: Comparison of experiment with theory and simulation models.
\newblock {\em Physics of Fluids}, 33(3), 2021.

\bibitem{jamali2019alternative}
Safa Jamali and John~F Brady.
\newblock Alternative frictional model for discontinuous shear thickening of dense suspensions: Hydrodynamics.
\newblock {\em Physical Review Letters}, 123(13):138002, 2019.

\bibitem{wang2020hydrodynamic}
Mu~Wang, Safa Jamali, and John~F Brady.
\newblock A hydrodynamic model for discontinuous shear-thickening in dense suspensions.
\newblock {\em Journal of Rheology}, 64(2):379--394, 2020.

\bibitem{guy2015towards}
Ben~M Guy, Michiel Hermes, and Wilson~CK Poon.
\newblock Towards a unified description of the rheology of hard-particle suspensions.
\newblock {\em Physical Review Letters}, 115(8):088304, 2015.

\bibitem{krishnamurthy2005shear}
Lakshmi-Narasimhan Krishnamurthy, Norman~J Wagner, and Jan Mewis.
\newblock Shear thickening in polymer stabilized colloidal dispersions.
\newblock {\em Journal of Rheology}, 49(6):1347--1360, 2005.

\bibitem{singh2022stress}
Abhinendra Singh, Grayson~L Jackson, Michael van~der Naald, Juan~J de~Pablo, and Heinrich~M Jaeger.
\newblock Stress-activated constraints in dense suspension rheology.
\newblock {\em Physical Review Fluids}, 7(5):054302, 2022.

\bibitem{raviv2001shear}
Uri Raviv, Rafael Tadmor, and Jacob Klein.
\newblock Shear and frictional interactions between adsorbed polymer layers in a good solvent.
\newblock {\em Journal of Physical Chemistry B}, 105(34):8125--8134, 2001.

\bibitem{james2018interparticle}
Nicole~M James, Endao Han, Ricardo Arturo~Lopez de~la Cruz, Justin Jureller, and Heinrich~M Jaeger.
\newblock Interparticle hydrogen bonding can elicit shear jamming in dense suspensions.
\newblock {\em Nature Materials}, 17(11):965--970, 2018.

\bibitem{jackson2022designing}
Grayson~L Jackson, Joseph~M Dennis, Neil~D Dolinski, Michael van~der Naald, Hojin Kim, Christopher Eom, Stuart~J Rowan, and Heinrich~M Jaeger.
\newblock Designing stress-adaptive dense suspensions using dynamic covalent chemistry.
\newblock {\em Macromolecules}, 55(15):6453--6461, 2022.

\bibitem{kimdynamic2023}
Hojin Kim, Mike van~der Naald, Neil~D. Dolinski, Stuart~J. Rowan, and Heinrich~M. Jaeger.
\newblock Dynamic-bond-induced sticky friction tailors non-newtonian rheology.
\newblock {\em Soft Matter}, 19:6797--6804, 2023.

\bibitem{brown2010generality}
Eric Brown, Nicole~A Forman, Carlos~S Orellana, Hanjun Zhang, Benjamin~W Maynor, Douglas~E Betts, Joseph~M DeSimone, and Heinrich~M Jaeger.
\newblock Generality of shear thickening in dense suspensions.
\newblock {\em Nature Materials}, 9(3):220--224, 2010.

\bibitem{raghavan1997shear}
Srinivasa~R Raghavan and Saad~A Khan.
\newblock Shear-thickening response of fumed silica suspensions under steady and oscillatory shear.
\newblock {\em Journal of Colloid and Interface Science}, 185(1):57--67, 1997.

\bibitem{jamali2015microstructure}
Safa Jamali, Arman Boromand, Norman Wagner, and Joao Maia.
\newblock Microstructure and rheology of soft to rigid shear-thickening colloidal suspensions.
\newblock {\em Journal of Rheology}, 59(6):1377--1395, 2015.

\bibitem{mewis2000rheology}
Jan Mewis and Jan Vermant.
\newblock Rheology of sterically stabilized dispersions and latices.
\newblock {\em Progress in Organic Coatings}, 40(1-4):111--117, 2000.

\bibitem{frith1996shear}
William~J Frith, P~d’Haene, R~Buscall, and Joannes Mewis.
\newblock Shear thickening in model suspensions of sterically stabilized particles.
\newblock {\em Journal of Rheology}, 40(4):531--548, 1996.

\bibitem{strivens1976shear}
TA~Strivens.
\newblock The shear thickening effect in concentrated dispersion systems.
\newblock {\em Journal of Colloid and Interface Science}, 57(3):476--487, 1976.

\bibitem{smith2010dilatancy}
MI~Smith, R~Besseling, ME~Cates, and V~Bertola.
\newblock Dilatancy in the flow and fracture of stretched colloidal suspensions.
\newblock {\em Nature Communications}, 1(1):114, 2010.

\bibitem{Isa2008}
L~Isa.
\newblock {\em Capillary flow of dense colloidal suspensions}.
\newblock PhD thesis, University of Edinburgh, 2008.

\bibitem{d1993scattering}
P~d'Haene, Joannes Mewis, and GG~Fuller.
\newblock Scattering dichroism measurements of flow-induced structure of a shear thickening suspension.
\newblock {\em Journal of Colloid and Interface Science}, 156(2):350--358, 1993.

\bibitem{foss2000structure}
David~R Foss and John~F Brady.
\newblock Structure, diffusion and rheology of brownian suspensions by stokesian dynamics simulation.
\newblock {\em Journal of Fluid Mechanics}, 407:167--200, 2000.

\bibitem{park2019contact}
Nayoung Park, Vikram Rathee, Daniel~L Blair, and Jacinta~C Conrad.
\newblock Contact networks enhance shear thickening in attractive colloid-polymer mixtures.
\newblock {\em Physical Review Letters}, 122(22):228003, 2019.

\bibitem{mari2014shear}
Romain Mari, Ryohei Seto, Jeffrey~F Morris, and Morton~M Denn.
\newblock Shear thickening, frictionless and frictional rheologies in non-brownian suspensions.
\newblock {\em Journal of Rheology}, 58(6):1693--1724, 2014.

\bibitem{andreotti2012shear}
Bruno Andreotti, Jean-Louis Barrat, and Claus Heussinger.
\newblock Shear flow of non-brownian suspensions close to jamming.
\newblock {\em Physical Review Letters}, 109(10):105901, 2012.

\bibitem{lerner2012unified}
Edan Lerner, Gustavo D{\"u}ring, and Matthieu Wyart.
\newblock A unified framework for non-brownian suspension flows and soft amorphous solids.
\newblock {\em Proceedings of the National Academy of Sciences}, 109(13):4798--4803, 2012.

\bibitem{fredrickson1991drainage}
Glenn~H Fredrickson and P~Pincus.
\newblock Drainage of compressed polymer layers: dynamics of a" squeezed sponge".
\newblock {\em Langmuir}, 7(4):786--795, 1991.

\bibitem{meng2006telechelic}
Xiao-Xia Meng and William~B Russel.
\newblock Telechelic associative polymers: Interaction potential and high frequency modulus.
\newblock {\em Journal of Rheology}, 50(2):169--187, 2006.

\bibitem{singh2018constitutive}
Abhinendra Singh, Romain Mari, Morton~M Denn, and Jeffrey~F Morris.
\newblock A constitutive model for simple shear of dense frictional suspensions.
\newblock {\em Journal of Rheology}, 62(2):457--468, 2018.    
\end{thebibliography}

\newpage

\begin{thebibliography}{10}

\bibitem{crucho2017functional}
Carina~IC Crucho, Carlos Baleiz\~ao, and Jos\'e Paulo~S Farinha.
\newblock Functional group coverage and conversion quantification in nanostructured silica by \textsuperscript{1}\uppercase{H} \uppercase{NMR}.
\newblock {\em Analytical Chemistry}, 89(1):681--687, 2017.

\end{thebibliography}
\end{document}